\documentstyle[aps,epsf]{revtex} 
\begin{document} 
\draft 
\title{Random Heteropolymer Dynamics} 
\author{Z. Konkoli$^{1}$, J. Hertz$^{1}$ and S. Franz$^{2}$} 
\address{ 
  $^1${NORDITA, Blegdamsvej 17, DK 2100 K\o benhavn, Denmark}\\ 
  $^2${The Abdus Salam ICTP, 
Strada Costiera 11, 
P.O. Box 563, 
34100 Trieste, Italy }} 
 
\date{\today} 
\maketitle 
\begin{abstract} 
 
We study the Langevin dynamics of the standard random heteropolymer 
model by mapping the problem to a supersymmetric field theory using 
the Martin-Siggia-Rose formalism. The resulting model is solved 
non-perturbatively employing a Gaussian variational approach. In 
constructing the solution, we assume that the chain is very long and 
impose the translational invariance which is expected to be present in 
the bulk of the globule by averaging over the center the of mass 
coordinate.  In this way we derive equations of motion for the 
correlation and response functions $C(t,t')$ and $R(t,t')$. The order 
parameters are extracted from the asymptotic behavior of these 
functions.  We find a dynamical phase diagram with frozen (glassy) and 
melted (ergodic) phases.  In the glassy phase the system fails to 
reach equilibrium and exhibits aging of the type found in p-spin 
glasses.  Within the approximations used in this study, the random 
heteropolymer model can be mapped to the problem of a manifold in a 
random potential with power law correlations. 
 
\end{abstract} 

\section{Introduction} 
 
Disordered systems can be extremely hard to solve, as the example of 
spin glasses shows \cite{SpGl}. It took enormous effort to understand 
the physics of infinite-dimensional spin glasses, while the fate of 
finite-dimensional spin glasses is still debated. Certainly, the 
complexity of the spin glass energy landscape is the major obstacle 
one has to deal with, and there are other systems sharing this 
feature: standard examples are proteins \cite{WE} and manifolds in 
random potentials \cite{NR}. 
 
The study of simplified random heteropolymer models may provide a 
useful first step toward understanding the physics of proteins.  Here, 
a central question is whether the trapping of the protein in a valley 
of the rough energy landscape can hinder, or perhaps even prevent, 
folding into its native state.  Something related to this scenario has 
actually been observed in some real proteins: the protein can be 
heated and then, upon re-cooling, misfold and never be able to find 
its native state \cite{Wol1,Wol2}. 
 
Here, we analyze the kind of dynamical trapping that can occur in the 
standard model of random heteropolymer \cite{garel,SG1}. 
 
So far, in addition to numerical simulations, two analytic approaches 
have been used to solve such models: equilibrium analysis employing 
the replica technique (see, e.g., refs. 
\cite{SG1,SG2,GHLO,SW,GLO,TW,GOP}) and dynamical studies using 
Langevin dynamics \cite{RS,TAB,TPW,Olem1,Pit,LT,Olem2,PS}. 
 
In the equilibrium approach, one studies the properties of Gibbs 
equilibrium.  Even the simplest kind of random heteropolymer model can 
only be approached analytically in approximated ways. In 
ref. \cite{SG1,SG2,GHLO,SW,GLO,TW,GOP} the model was analyzed with 
replica variational approximations, which predict ergodicity breaking 
at low temperature, giving one-step replica symmetry breaking (1RSB) 
for $d>2$ and continuous replica symmetry breaking for $d<2$. Thus, in 
3D, below the freezing temperature, the ergodic components lie far 
apart from each other, and the same interstate overlap $q_0$. Given 
the intrinsic one dimensional nature of the polymer, it has been 
argued that RSB could be an artifact of the variational 
approximation.\cite{elba} We believe however, that even in this 
eventuality, implying that a single native state dominate the 
thermodynamics, RSB in the variational approximation is a signal of a 
complex energy landscape, which can lead to slow dynamics, with 
off-equilibrium behavior on long time scales. 
 
Models exhibiting 1RSB (such as the simple random heteropolymer 
mentioned above, the p-spin glass, or a manifold in a random 
potential) have been found to have different dynamic and static phase 
diagrams, with a dynamical energy density higher then the one found at 
equilibrium.  This raises the intriguing possibility that, for a 
suitable range of temperatures and times, a heteropolymer might find 
itself dynamically trapped in a local state (as in the scenario 
described above), while the equilibrium statistical mechanics might 
give no clue that this was happening.  Such trapping would thus be an 
intrinsically non-equilibrium effect, and a dynamical theory is 
required to describe it. 
 
For the models with this feature, the solutions exhibit a breakdown of 
time-translation invariance (the correlation functions depend on the 
time since the system was quenched into the glassy state) and a 
breakdown of the fluctuation-dissipation relation (which is a 
fundamental characteristic of Gibbs equilibrium).  Together, these 
properties of the dynamical glassy phase go under the name ``aging'', 
and it is one of our goals here to examine the possibility of aging in 
heteropolymers. 
 
In this paper we consider the simple random heteropolymer model with 
Langevin dynamics (as in ref. \cite{TPW,PS}). The equations of motion 
are constructed in such a way that the Gibbs distribution is the 
stationary solution of the dynamics. This type of dynamical approach 
was used successfully in spin glass models. 
 
To derive closed equations of motion for correlation and response 
functions we resort to a Gaussian variational ansatz similar to the 
one used at equilibrium. The same approach has been used to study the 
problem of a manifold in a random potential, for both statics 
\cite{MP1,MP2} and dynamics \cite{CD,CKD}. In related dynamical work 
on random heteropolymer model \cite{TPW} and \cite{PS}, the slightly 
different approach of Mode Coupling Theory is used. Our approach gives 
results fully coherent with the ones obtained there, although the 
detailed form of the phase diagram differs, due to the different 
nature of the approximation. 
 
The analysis of the variational equations indicates that, as expected 
from static treatments, the random heteropolymer model exhibits 
spontaneous breaking of ergodicity in a glassy phase.  All these 
states are equally distant from each other; they have same interstate 
overlap (naturally, the self overlap is different).  We also discuss 
the nature of the transition from the frozen (glassy) to the melted 
(ergodic) phase.  Furthermore, we find that, within the Gaussian 
variational approximation that we employ, the random heteropolymer 
model can be mapped onto the the problem of a manifold in a random 
potential with power law correlations.

The paper is organized as follows.  Section II briefly describes the 
Langevin model. In section III a mapping to a supersymmetric (SUSY) 
field theory is made. The resulting action can be simplified by 
assuming a very long chain. This is discussed in section IV. 
Dynamical equations in SUSY notation, given in section VII, are 
obtained via the variational ansatz discussed in section V and 
VI. Also, in section VII, the connection of the random heteropolymer 
model to the problem of a manifold in a random potential will be 
commented upon. After disentangling the SUSY notation, one obtains 
dynamical equations for correlation and response functions (section 
VIII).  An analytical ansatz for solving these equations is introduced 
in section IX, and the solution is obtained in section X.  Section XI 
discusses the ergodic phase, while in section XII the spin glass phase 
is analyzed. Technicalities needed to construct the full phase diagram 
are given in section XIII.

\section{The Model} 
 
The model is defined as follows. The Langevin dynamics is assumed to 
be governed by the Hamiltonian $H[x]$, 
\begin{equation} 
  \partial x(s,t)/\partial t = - \partial H[x] / \partial x(s,t) + \eta(s,t), 
  \label{eq:dxdt} 
\end{equation} 
where $x(s,t)$ is the position of chain bead $s$ at time $t$. Beads 
are numbered continuously from $s=0$ to $s=N$. $\eta(s,t)$ is Gaussian 
noise: 
\begin{equation} 
  \langle \eta(s,t)\eta(s',t') \rangle_T = 2\delta(s-s')\delta(t-t') T 
  \label{eq:etas} 
\end{equation} 
due to contact with a heat bath at temperature $T$. 
The Hamiltonian $H[x]=H_0[x]+H_{rand}[x]$ contains a deterministic 
part $H_0[x]$ and a random part $H_{rand}[x]$. The $H_0[x]$ is defined 
as 
\begin{equation} 
   H_0[x]= \frac{T}{2} \int_{0}^{N} ds  
        [(\partial x(s,t)/\partial s)^2+\mu x(s,t)^2] 
   \label{eq:H0} 
\end{equation} 
and describes the elastic properties of the chain and a confinement 
potential which fixes the density of the protein.  The random part 
$H_{rand}$ describes heterogeneity of the interactions between the 
beads, 
\begin{equation} 
   H_{rand}[x]= \frac{1}{2} \int_{0}^{N} ds ds'  
              B_{s,s'} V(x(s,t)-x(s',t)). 
   \label{eq:Hrand} 
\end{equation} 
$B_{s,s'}$ is quenched Gaussian noise with variance $B^2$: 
\begin{equation} 
  \langle B_{s,s'}^2 \rangle_B = B^2 , \ \ s>s' . 
\end{equation} 
$V(\Delta x)$ is a short-range potential, and for simplicity we 
take it to have a Gaussian form, as in ref. \cite{TPW}: 
\begin{equation} 
   V(\Delta x)=(\frac{1}{2\pi\sigma})^{d/2}e^{-(\Delta x)^2/2\sigma}. 
   \label{eq:V} 
\end{equation} 
$d$ is the dimensionality of the system, and $\sigma$ parameterizes 
the range of the potential. Large (small) $\sigma$ results in a long- 
(short-) range potential. In particular, for $\sigma\rightarrow 0$, 
$V(\Delta x)\rightarrow\delta(\Delta x)$, and we recover the potential 
used in \cite{PS}. Here and in the following $\Delta x$ denotes the 
bead-to-bead distance: $\Delta x=x(s,t)-x(s',t)$ for a pair of beads 
$s$, $s'$. 
 
This model admits a stationary solution characterized by a Gibbs 
distribution.  The equilibrium partition function for this solution is 
given by 
\begin{equation} 
  {\cal Z} = \int Dx  
      e^{  
         - \frac{1}{2} \int_{0}^{N} ds  
           [(\partial x(s)/\partial s)^2+\mu x(s)^2] 
         - \frac{\beta}{2} \int_{0}^{N} ds ds'  
              B_{s,s'} V(x(s)-x(s')) 
        }. 
  \label{eq:Z} 
\end{equation} 
$T$ in Eq.(\ref{eq:H0}) ensures that the chain constraint and 
quadratic confinement are temperature-independent.  (That is, the 
elasticity is purely entropic in origin.)  The same convention was 
used in ref. \cite{TPW}. This differs slightly from the work in ref 
\cite{PS}, where the elastic term had a factor $\beta$ in front of it. 
Our choice ensures that for high temperatures the random heteropolymer 
behaves as a Gaussian random coil. Also, for very low temperatures, 
the random part of interaction with $\beta$ in front dominates 
($\beta\rightarrow\infty$; the elastic and confinement terms become 
negligible). Thus, in principle, for $\beta=\infty$, ${\cal Z}$ in 
Eq.~(\ref{eq:Z}) is dominated by minima of $\int ds ds' B_{s,s'} 
V(x(s)-x(s'))$. Furthermore, in this limit there is nothing that would 
control the spatial spread of those minima, and $\langle x^2(s,t) 
\rangle_T$ diverges for very low temperatures. (This only happens when 
$\mu$ is held fixed.  If it it adjusted appropriately, one can keep 
$\langle x^2(s,t) \rangle_T$ fixed instead.  In this paper, however, 
we will be concerned with finite-$T$ phase transitions, not the 
low-$T$ limit, so we will work with fixed $\mu$.)

\section{Mapping to the Field Theory}

Using the standard Martin-Siggia-Rose formalism \cite{MSR}, the 
dynamical average of any observable can be calculated as 
\begin{equation} 
  \langle {\cal O}[x,\tilde x] \rangle_T =\int Dx D\tilde x D D\xi D\bar\xi  
  {\cal O}(x,\tilde x) e^{-S[x,\tilde x,\xi,\bar\xi] }, 
  \label{eq:average} 
\end{equation} 
with the following dynamical action: 
\begin{eqnarray} 
  S[x,\tilde x,\xi,\bar\xi] = && 
     \int dt ds 
     \left[ 
          - T \tilde x(s,t)^2  
          + \tilde x(s,t) \left(  
              \frac{\partial}{\partial t} x(s,t)  
              + \frac{\partial H[x]}{\partial x(s,t)} 
            \right)  
     \right] \cr  
     && - \int dt ds \bar\xi(s,t)\frac{\partial}{\partial t}\xi(s,t) 
     + \int dt ds ds' \bar\xi(s,t) 
         \frac{\partial^2 H[x]}{\partial x(s,t) \partial x(s',t)} \xi(s',t) 
  \label{eq:S}  
\end{eqnarray}  
$\tilde x$, $\xi$, $\bar\xi$ are auxiliary fields which appear in the 
formalism.  To compactify the notation we introduce the superfield $\Phi$: 
\begin{equation} 
  \Phi(s,t_1,\theta_1,\bar\theta_1)=x(s,t_1)+\bar\xi(s,t_1)\theta_1  
  + \bar\theta_1\xi(s,t_1) + \bar\theta_1\theta_1\tilde x(s,t_1), 
  \label{eq:Phi} 
\end{equation} 
where $\theta$ and $\bar\theta$ are Grassmann (anti-commuting) 
variables. For $X,X' \in \{\theta, \bar\theta,\theta',\bar\theta'\}$, 
$\{X,X'\}=0$ and $\int dX X=1$, the rest of the integrals being 
zero. In the following, for practical reasons, the more compact 
notation $\Phi(s,1)\equiv\Phi(s,t_1,\theta_1,\bar\theta_1)$ will be 
used. Also, the integral symbol $\int d\theta_1 d\bar\theta_1 dt_1$ 
will be denoted by $\int d1$. 
 
In supersymmetric (SUSY) notation Eqs.~(\ref{eq:average}) and (\ref{eq:S})  
translate into (\ref{eq:avSUSY}) and (\ref{eq:SSUSY}): 
\begin{eqnarray} 
  && \langle {\cal O}[\Phi] \rangle_T =\int D\Phi  
    {\cal O}[\Phi] e^{-S[\Phi] },           \label{eq:avSUSY} \\ 
  && S[\Phi] = S_0[\Phi]+S_{rand}[\Phi],      \label{eq:SSUSY}  
\end{eqnarray}  
where 
\begin{eqnarray} 
  && S_0[\Phi]= 1/2 \int ds d1 ds' d2 \Phi(s,1) K_{12}^{ss'} \Phi(s'2), 
  \label{eq:S0} \\ 
  && S_{rand}[\Phi]= 1/2 \int d1 ds ds' B_{s,s'} V(\Phi(s,1)-\Phi(s',1)), 
  \label{eq:Srand} 
\end{eqnarray} 
and 
\begin{eqnarray} 
  K_{12}^{ss'} && \equiv \delta_{12} \delta_{ss'} K_1^s \ , \ \ 
  K_1^s  = T \left[ \mu-(\partial/\partial s)^2 \right] - D_1^{(2)}, \\  
  D_1^{(2)} && =2 T \frac{\partial^2}{\partial\theta_1\partial\bar\theta_1} + 
   2 \theta_1 \frac{\partial^2}{\partial\theta_1\partial t_1} -  
  \frac{\partial}{\partial t_1}, 
\end{eqnarray} 
As noticed by De Dominicis \cite{Dom} the expression in 
Eq.(\ref{eq:avSUSY}) is already normalized, so the average over the 
quenched random interactions $B_{s,s'}$ can be done directly on 
(\ref{eq:avSUSY}): 
\begin{equation} 
  \langle\langle A[\Phi] \rangle_T\rangle_B = \int D\Phi 
  A[\Phi] e^{-(S_0[\Phi]+S_1[\Phi])}, 
  \label{eq:avA} 
\end{equation} 
where $\exp(-S_1[\Phi])\equiv\langle\exp(-S_{rand}[\Phi])\rangle_B$. 
The average over $B_{s,s'}$ can be done easily, leading to 
\begin{equation} 
  S_1[\Phi] = -B^2/4 \int ds ds'  
                  \left[  
                      \int d1 V(\Phi(s,1)-\Phi(s',1)) 
                  \right]^2. 
\end{equation} 
The dynamical action $S=S_0+S_1$ closely resembles the effective 
Hamiltonian obtained in the static replica approach of 
ref. \cite{SG1,SG2}.  (This rather general similarity between replica 
and SUSY treatments has been discussed in ref. \cite{Kur}.)  Instead 
of summation over replica indices in \cite{SG1,SG2} we have $\int 
d1$. Our expressions are not identical to those in \cite{SG1,SG2}, 
since we use a quadratic well potential instead of two- and three-body 
interaction terms to confine the polymer.  Also, we use a Gaussian 
$V(\Delta x)$ instead of $\delta(\Delta x)$.

\section{Long Chain approximation} 
 
The $S_1$ part of the action can be further simplified. It can be 
rewritten in the form 
\begin{equation} 
  S_1 = -\frac{B^2}{4} A^{(V)} * A^{(\delta)}. 
\end{equation} 
with the notation 
\begin{equation} 
  A^{(V)} * A^{(\delta)} = \int d1 d2 dx dy  
     A^{(V)}_{1,2}(x,y) A^{(\delta)}_{1,2}(x,y), 
  \label{av*ad} 
\end{equation} 
where $A^{(V)}$ and $A^{(\delta)}$ are given by 
\begin{eqnarray} 
  &&  A^{(V)}_{1,2}(x,y) =  
               \int ds V(\Phi(s,1)-x) V(\Phi(s,2)-y) \label{av} \\ 
  &&  A^{(\delta)}_{1,2}(x,y) =  
               \int ds' \delta(\Phi(s',1)-x)  
                  \delta(\Phi(s',2)-y).               \label{ad}  
\end{eqnarray} 
It is useful to transform $\exp(-S_1)$ as 
\begin{eqnarray} 
   && \exp\left[  
         \frac{B^2}{4} A^{(V)} * A^{(\delta)}  
      \right]    
    = \exp 
      \left[  
         \frac{B^2}{16}  
            \left[  
               (A^{(V)}+A^{(\delta)})*(A^{(V)}+A^{(\delta)}) 
              -(A^{(V)}-A^{(\delta)})*(A^{(V)}-A^{(\delta)}) 
            \right]  
      \right] \nonumber \\  
   &&  = \int DQ_1 DQ_2 \exp\left[ \frac{B^2}{4}  
      \left[-(Q_1*Q_1+Q_2*Q_2)+Q_1 * (A^{(V)}+A^{(\delta)})  
      + i Q_2 * (A^{(V)}-A^{(\delta)})\right] \right].   
   \label{eq:expS1} 
\end{eqnarray} 
Then, the dynamical generating functional $F$ 
defined by 
\begin{equation} 
  e^{-F}=\int D\Phi e^{-S[\Phi]+J*\Phi}, 
\end{equation} 
with $J*\Phi=\int ds d1 J(s,1) \Phi(s,1)$, can be written as 
\begin{equation} 
  e^{-F} = \int DQ_1 DQ_2 e^{- \frac{B^2}{4} ( Q_1*Q_1 + Q_2*Q_2)} 
              \int D \Phi e^{L[Q_1,Q_2,\Phi]}, 
  \label{eq:FJ} 
\end{equation} 
with $L$ given by 
\begin{equation} 
  L = \frac{B^2}{4} [Q_1 * (A^{(V)}+A^{(\delta)})  
                     + i Q_2 * (A^{(V)}-A^{(\delta)})]  
              - S_0[\Phi] + J*\Phi . 
  \label{eq:L}   
\end{equation} 
So far everything was exact. $A^{(V)}$ and $A^{(\delta)}$ are both of 
order $N$ and for very long chains one can calculate integrals over 
$Q_1$ and $Q_2$ in (\ref{eq:FJ}) using a saddle point 
approximation. The saddle point equations read 
\begin{eqnarray} 
  && Q_1^{s.p.} = \frac{1}{2} \langle A^{(V)}+A^{(\delta)} \rangle_{L'}  \\ 
  && Q_2^{s.p.} = \frac{i}{2} \langle A^{(V)}-A^{(\delta)} \rangle_{L'}, 
  \label{eq:Q12} 
\end{eqnarray} 
where $\langle \rangle_{L'}$ denotes the average with $L$ taking 
$Q_1$, $Q_2 \rightarrow Q_1^{s.p.}, Q_2^{s.p.}$. This leads to 
self-consistent equations for $Q_1^{s.p.}$ and $Q_2^{s.p.}$. 
 
Thus, Eq.(\ref{eq:FJ}) can be approximated as 
\begin{equation} 
  e^{-F} \approx \int D\Phi e^{-S'[\Phi]+J*\Phi}  
  \label{eq:FJ'}, 
\end{equation} 
with $S'[\Phi]=S_1'[\Phi]+S_0[\Phi]$ and 
\begin{equation} 
  S_1'[\Phi] = \frac{B^2}{4}  
   \left[  
      \langle A^{(V)} \rangle_{S'} * \langle A^{(\delta)} \rangle_{S'} 
      - A^{(V)} * \langle A^{(\delta)} \rangle_{S'}  
      - A^{(\delta)} * \langle A^{(V)} \rangle_{S'}  
    \right]. 
  \label{eq:S'} 
\end{equation} 
$\langle A^{(V)} \rangle_{S'}$ and $\langle A^{(\delta)} \rangle_{S'}$ 
have to be calculated self consistently with $S'$: 
\begin{equation} 
  \langle A^{(V,\delta)} \rangle_{S'} =  
     \frac{\int D\Phi A^{(V,\delta)}  
        e^{-S'+J*\Phi}}{\int D\Phi e^{-S'+J*\Phi}},  
     \label{eq:Avd} \\ 
\end{equation} 
In the limit $N\rightarrow\infty$ 
Eq.s(\ref{eq:FJ'},\ref{eq:S'},\ref{eq:Avd}) provide an exact 
description of random heteropolymer dynamics.

\section{Variational ansatz} 
 
To solve the model we proceed by using a variational ansatz, assuming 
that fields $\Phi$ are approximately described by a Gaussian action 
\begin{equation} 
  S_{var}=\frac{1}{2} \int d1 ds d2 ds'  
    \Phi(s,1) G(s,1;s',2)^{-1} \Phi(s',2).  
\end{equation} 
This approach has been widely used in statics.  Here we apply it to a 
dynamic calculation.  The goal is to calculate $F$ given by 
Eq.(\ref{eq:FJ'}). Since the variational parameter $G(s,1;s',2)$ is 
the only quantity we are interested in, there is no need to keep the 
source $J$. It is convenient to write Eq.~(\ref{eq:FJ'}), with $J=0$, 
formally as 
\begin{equation} 
  e^{-F} = \langle e^{-(S'-S_{var})}  
  \rangle_{var} e^{-F_{var}} 
  \label{eq:FJvar}, 
\end{equation} 
where 
\begin{equation} 
  e^{-F_{var}}=\int D\Phi e^{-S_{var}}$\ ,\ \ \ \  $\langle 
  . \rangle_{var}=e^{F_{var}}\int D\Phi ( . ) e^{-S_{var}}. 
\end{equation} 
In usual statics, for problems without disorder, the variational 
approach is related to a maximum principle.  The equivalent of 
Eq.(\ref{eq:FJvar}) leads to the inequality 
\begin{equation} 
   e^{-F} \ge e^{  - \langle (S'-S_{var}) \rangle_{var} } 
              e^{ -F_{var} }. 
   \label{eq:FJFdyn} 
\end{equation} 
In the present dynamical problem, as well as in the static problem 
with replicas, unfortunately such a maximum principle is not known, 
and the variational free-energy cannot be claimed to be an upper bound 
to the true one. Despite that, the variational approach has been 
argued to give exact results in some limiting cases \cite{MP1,MP2}, 
giving a justification for its use in general. 
 
The dynamical variational free-energy $F_{dyn}=\langle (S'-S_{var}) 
\rangle_{var}+F_{var} $ is given by 
\begin{equation} 
F_{dyn} = F_{dyn}^{(1)} + F_{dyn}^{(2)} + F_{dyn}^{(3)}, 
\label{eq:Fdyn} 
\end{equation} 
with 
\begin{eqnarray} 
  && F_{dyn}^{(1)} = \frac{d}{2} \int ds d1 ds' d2  
     K_{12}^{ss'} G_{12}^{ss'}  \label{eq:Fdyn1} \\  
  && F_{dyn}^{(2)} = - \frac{d}{2} Tr \ln G \label{eq:Fdyn2} \\ 
  && F_{dyn}^{(3)} = - \frac{B^2}{4} \langle A^{(V)} \rangle_{var} *  
                    \langle A^{(\delta)} \rangle_{var}. \label{eq:Fdyn3} 
\end{eqnarray} 
Note that in calculating $F_{dyn}^{(3)}$, the average $\langle\ 
\rangle_{S'}$ in Eq.(\ref{eq:S'}) is performed over the trial 
distribution (and therefore denoted $\langle\ \rangle_{var}$).

\section{ Calculating  
$\langle A^{(V)} \rangle * \langle A^{(\delta)} \rangle$}

The term $\langle A^{(V)} \rangle_{var} * \langle A^{(\delta)} 
\rangle_{var}$ is the only non-trivial term in $F_{dyn}$. Before 
calculating it we simplify each of the factors in product further 
\begin{equation} 
  A^{(V,\delta)}_{1,2}(x,y) \approx A^{(V,\delta)}_{1,2}(u)  
   \equiv 1/v \int d^dR \, A^{(V,\delta)}_{1,2}(u,R), 
   \label{eq:Avd_av} 
\end{equation} 
where $v$ is the volume of the protein and a new coordinate system has 
been introduced: 
\begin{equation} 
  R=(x+y)/2, u=(x-y)/2.  
\end{equation} 
$R$ is the center of mass and $u$ a relative distance coordinate. 
Thus, translational invariance is introduced by hand via 
Eq.(\ref{eq:Avd_av}).  This approximation is not necessary; the model 
could be solved without it. However, as we shall see later on, this 
approximation leads to dynamical equations which are identical to 
those for the random manifold model studied in \cite{CD,CKD,FM}. 
 
Changing integration variables from $d^dx\, d^dy$ to $d^dR\, d^du$ 
(the Jacobian is $2^d$) gives 
\begin{equation} 
  \int d1\, d2\, d^dx\, d^dy  
    \langle A^{(V)}_{1,2}(x,y) \rangle_{var} 
    \langle A^{(\delta)}_{1,2}(x,y) \rangle_{var} 
  \approx 
    \frac{2^d}{v}  
    \int d1\, d2\, d^du  
     \int d^dR' A^{(V)}_{1,2}(u,R')  
     \int d^dR'' A^{(\delta)}_{1,2}(u,R'')  
\end{equation} 
The integrals over $R'$ and $R''$ can be easily performed and one gets 
\begin{equation} 
  \frac{2^d}{v}  
    \int d1\, d2\, d^du\, ds\, ds'\, d^d\alpha \, d^d\beta \,   
    V(\alpha) \, V(\beta)  
    \langle  
       \delta \left[ 2 u - \alpha + \beta - \Phi(s,1) + \Phi(s,2) \right] 
    \rangle_{var} 
    \langle  
       \delta \left[ 2 u - \Phi(s,1) + \Phi(s,2) \right] 
    \rangle_{var}, 
\end{equation} 
which can be further written as 
\begin{equation} 
  \frac{2^d}{v} 
    \int d1\, d2\, d^du\, ds\, ds'\, d^d\alpha\, d^d\beta \, 
    V(\alpha) \, V(\beta)  
    \int \frac{d^d p}{(2\pi)^d} \frac{d^d q}{(2\pi)^d}  
     e^{i(p+q)2u} e^{ip(\beta-\alpha)}  
    e^{-\frac{1}{2}p^2 B_{12}^s}  
    e^{-\frac{1}{2} q^2 B_{12}^{s'}},  
   \label{eq:Avd_av1} 
\end{equation} 
where averages over $S_{var}$ have been evaluated as 
\begin{equation} 
   \langle e^{ip(\Phi(s,1)-\Phi(s,2))} \rangle_{var} = 
    e^{-\frac{1}{2}p^2 B_{12}^s},  
\end{equation} 
with $B_{1,2}^s$ is given by 
\begin{equation} 
  B_{1,2}^s = \langle \left[\Phi(s,1)-\Phi(s,2)\right]^2 \rangle_{var} 
          = G(s,1;s,1) + G(s,2;s,2) - 2 G(s,1;s,2) 
\end{equation} 
Integrating Eq.(\ref{eq:Avd_av1}) first over $u$,  
and then over $q$ and $p$ finally gives 
\begin{equation} 
  \frac{1}{v} 
    (2\pi)^{-d/2} \int d1 d2 ds ds' d^d\alpha d^d\beta 
    V(\alpha) V(\beta) (B_{12}^s+B_{12}^{s'})^{-d/2} 
    \exp{- \frac{ 
           (\alpha-\beta)^2  
              }{  
           2 (B_{12}^s+B_{12}^{s'})  
         }  
      }. 
   \label{eq:AvdCM} 
\end{equation} 
Eq.(\ref{eq:AvdCM}) holds for any $V(\Delta x)$. However, technically, 
it is of little use unless the integrals over $\alpha$ and $\beta$ can 
be performed explicitly. For the Gaussian form for $V(\Delta x)$ (see 
Eq.\ref{eq:V}) it is possible to perform the integrals, and one gets 
\begin{equation} 
      \langle A^{(V)}\rangle_{var} * \langle A^{(\delta)} \rangle_{var}  
    \approx 
      \frac{1}{v}  (4\pi)^{-d/2} \int d1 d2 ds ds'  
      \left[ (B_{12}^s+B_{12}^{s'})/2 + \sigma \right]^{-d/2} 
  \label{eq:V2} 
\end{equation} 
and finally 
\begin{equation} 
  F_{dyn}^{(3)} \approx \frac{d}{2N} \int d1 d2 ds ds'  
      {\cal V} \left[ (B_{12}^s+B_{12}^{s'})/2 \right] 
  \label{eq:Fdyn3V} 
\end{equation} 
with 
\begin{equation} 
  {\cal V}(z) = - \frac{\tilde B^2}{d} (z+\sigma)^{-d/2}\ ,\ \   
  \tilde B^2 = \frac{B^2}{2} \frac{N}{v} (4\pi)^{-d/2}. 
  \label{eq:Vz} 
\end{equation} 
Eq.s (\ref{eq:Fdyn},\ref{eq:Fdyn1},\ref{eq:Fdyn2},\ref{eq:Fdyn3}) and  
(\ref{eq:Fdyn3V}) fully determine $F_{dyn}$.

\section{Equations of motion in SUSY notation}

Given the $F_{dyn}$,  one can derive the equations of motion 
from the stationarity condition 
\begin{equation} 
  \frac{\delta}{\delta G_{12}^{ss'}} F_{dyn} = 0.  
  \label{eq:dFdyn} 
\end{equation} 
The most complicated term  is $\frac{\delta}{\delta 
G_{12}^{ss'}} F_{dyn}^{(3)}$.  From (\ref{eq:Fdyn3V}), it is   
\begin{equation} 
  \frac{d}{2N} 
  \int d3 d4 du dv  
     {\cal V}' \left[ ( B_{34}^u+B_{34}^v )/2 \right]  
     \delta_{ss'} \frac{\delta_{us}+\delta_{vs}}{2} 
     ( \delta_{13}\delta_{23} + \delta_{14}\delta_{24} -  
       \delta_{13}\delta_{24} - \delta_{14}\delta_{23} ). 
  \label{eq:dFdyn3V} 
\end{equation} 
Due to translational invariance in $s$, $B_{12}^s$ is independent of 
$s$.  After dropping the index $s$ Eq.(\ref{eq:dFdyn3V}) simplifies to 
\begin{equation} 
    \frac{\delta}{\delta G_{12}^{ss'}} F_{dyn}^{(3)} =  
    d \, \delta_{ss'}  
      \left[  
        \delta_{12} \int d3 {\cal V}'(B_{13}) - {\cal V}'(B_{12})  
      \right]. 
\end{equation} 
The variations of $F_{dyn}^{(1)}$ and $F_{dyn}^{(2)}$ are trivial. 
Using Eq.~(\ref{eq:dFdyn}) and (\ref{eq:Fdyn}) leads to 
\begin{equation} 
  K_{12}^{ss'} - ( G_{12}^{ss'} )^{-1} 
  + 2 \, \delta_{ss'} 
    \left[  
        \delta_{12} \int d3 {\cal V}(B_{13}) - {\cal V}(B_{12})  
    \right] = 0, 
  \label{eq:KGinv} 
\end{equation} 
which can be written as 
\begin{equation} 
  K_1^s G_{12}^{ss'} = \delta_{12} \delta_{ss'}  
   + 2 \int d3 {\cal V}'(B_{13})(G_{32}^{ss'}-G_{12}^{ss'}). 
  \label{eq:emSUSY} 
\end{equation} 
Due to translational invariance in the variable $s$ it is useful to 
define following Fourier transforms 
\begin{equation} 
  G_{12}^{ss'} \equiv \int_{-\infty}^{\infty} \frac{dk}{2\pi} 
  e^{ik(s-s')} G_{12}^k. 
\end{equation} 
Then Eq.(\ref{eq:emSUSY}) translates into 
\begin{equation} 
   \left[ T(\mu+k^2)-D^{(2)}_1 \right] G_{12}^k = \delta_{12} 
    + 2 \int d3 {\cal V}'(B_{13})(G_{32}^k-G_{12}^k).  
  \label{eq:emSUSYk} 
\end{equation} 
Eq.~(\ref{eq:emSUSYk}) is identical to the one obtained in 
ref. \cite{CKD} for a $D$-dimensional manifold $\varphi(\omega)$ 
($\omega\in R^D$, $\varphi\in R^d$) in a random potential 
$V(\varphi(\omega),\omega)$, where the correlations of the potential 
are described by 
\begin{equation} 
  \langle V(\varphi,\omega)V(\varphi',\omega') \rangle = - 
   d\delta^D(\omega-\omega') \hat{\cal V}\left[(\varphi-\varphi')/d\right]. 
\end{equation} 
These equations of motion were derived using the Gaussian Variational 
Approximation (GVA), which is exact for the random manifold problem in 
$d=\infty$. We expect the same behavior for the random 
heteropolymer. However, in this study we work at finite $d$, so the 
equations of motion are approximate. 
 
There have been several studies of random manifolds where $\hat{\cal 
V}$ describes power law correlations as in Eq.~(\ref{eq:Vhat}), 
employing static \cite{MP1,MP2,Eng}, and dynamical \cite{FM,KH1,KH2} 
approaches: 
\begin{equation} 
  \hat{\cal V}(z)=(z+\sigma)^{1-\gamma}/2(1-\gamma). 
  \label{eq:Vhat} 
\end{equation} 
By comparing Eq.s~(\ref{eq:Vz}) and (\ref{eq:Vhat}) one notices that 
${\cal V}(z)$ is identical to $\hat{\cal V}(z)$ (up to a 
proportionality factor $\tilde B^2$) if one identifies $\gamma=1+d/2$. 
Accordingly, we conclude that, within the Gaussian variational 
approximation used in this study, random heteropolymer dynamics is 
identical to the dynamics of the manifold in a random potential with 
power law correlations.  (We can not say anything rigorous outside the 
framework of the Gaussian variational approximation scheme, of 
course.) 
 
Furthermore, correlations of the random manifold potential are 
classified as short range for $\gamma > 2/(2-D)$, and long range for 
$\gamma < 2/(2-D)$ \cite{MP1,MP2}. This classification of random 
manifolds helps to classify random heteropolymer model in the same 
way. Using $\gamma=1+d/2$, the random heteropolymer has $D=1$, and 
short range correlations for $d>2$ and long range correlations for 
$d<2$. (Again, this all makes sense only within the Gaussian 
variational approximation).

\section{Disentangling SUSY}

$G_{12}^{ss'}$ encodes 16 correlation functions, out of which only 
two, correlation and response function, are independent and nonzero:  
\begin{eqnarray} 
  && \langle \langle x(s,t_1) x(s',t_2) \rangle \rangle  
     \equiv C(s,t_1;s',t_2) = 
     \int \frac{dk}{2\pi} e^{ik(s-s')}C_k(t_1,t_2) \label{eq:C} \\ 
  && \langle \langle x(s,t_1) \tilde x(s',t_2) \rangle \rangle  
     \equiv R(s,t_1;s',t_2) = 
     \int \frac{dk}{2\pi} e^{ik(s-s')} R_k(t_1,t_2). \label{eq:R} 
\end{eqnarray} 
Also, by adding an external field term to the original Hamiltonian $H[x] 
\rightarrow H[x]+\int ds dt x(s,t) h(s,t)$ one gets 
\begin{equation} 
  \langle\langle x(s,t_1) \tilde x(s',t_2) \rangle\rangle = 
   \frac{\delta}{\delta h(s',t_2)} \langle\langle x(s,t_1)\rangle\rangle. 
\end{equation} 
i.e. $R(s,t_1;s',t_2)$ describes the response to an infinitesimal 
field applied at time $t_2$ and bead $s'$. Thus, $G_{12}^k$ reduces to 
\begin{equation} 
  G_{12}^k = C_k(t_1,t_2) +  
    (\bar\theta_1-\bar\theta_2) \left[ \theta_1 R_k(t_2,t_1) -  
    \theta_2 R_k(t_1,t_2) \right], 
  \label{eq:G12k} 
\end{equation} 
and, accordingly, with $G_{11}^{ss}=C(s,t_1;s,t_1)$, 
$G_{22}^{ss}=C(s,t_2;s,t_2)$, and Eqs.~(\ref{eq:C},\ref{eq:R}), one 
gets 
\begin{equation} 
  B_{12} = B(t_1,t_2) - 2 (\bar\theta_1-\bar\theta_2)  
    \left[ \theta_1 r(t_2,t_1) - \theta_2 r(t_1,t_2) \right], 
\end{equation} 
with 
\begin{equation} 
   B(t_1,t_2) = \int \frac{dk}{2\pi}  
                    \left[ 
                       C_k(t_1,t_1)+C_k(t_2,t_2)-2 C_k(t_1,t_2) 
                    \right] 
\end{equation} 
and 
\begin{equation} 
   r(t_1,t_2) = \int \frac{dk}{2\pi} R_k(t_1,t_2).  
   \label{eq:r12} 
\end{equation} 
After disentangling the equations of motion in SUSY notation (see 
Eq.~\ref{eq:emSUSYk}) by using (\ref{eq:G12k}-\ref{eq:r12}) gives  
\begin{eqnarray} 
 & [T(\mu+k^2)+\partial/\partial t] C_k(t,t') = &  
  2 T R_k(t',t) + 
  2 \int_{0}^{t} ds {\cal V}'\left[B(t,s)\right] R_k(t',s) + \nonumber \\ 
  & & + 4 \int_{0}^{t} ds {\cal V}''\left[B(t,s)\right] r(t,s) \left[  
  C_k(t,t')-C_k(s,t') \right], \label{eq:emC} \\ 
  & [T(\mu+k^2)+\partial/\partial t] R_k(t,t') = & \delta(t-t') + 
  4 \int_{0}^{t} ds {\cal V}''\left[B(t,s)\right] r(t,s) \left[  
    R_k(t,t')-R_k(s,t') \right]. \label{eq:emR} 
\end{eqnarray} 
The equations of motion (\ref{eq:emC}) and (\ref{eq:emR}) are almost 
identical to the ones found in ref. \cite{CD} (here $D=1$, while in 
\cite{CD} $D=0$).

\section{Ansatz for $C_k(t,t')$ and $R_k(t,t')$ } 
\label{sec:ansatz} 
 
These equations of motion are coupled integro-differential equations 
which in principle can be solved; the initial conditions are given by 
$C_k(0,0)$ and we use Ito's convention $R(t+\epsilon,t)\rightarrow 1$ 
as $\epsilon\rightarrow 0$. It is well known that asymptotic solutions 
of such equations can be characterized by few parameters and it is 
possible to solve those equations 
analytically. \cite{CD,CKD,FM,CK1,BCKP,CK2} 
 
For $t,t'\rightarrow\infty$, $\frac{\tau}{t'}<<1$ and $\tau=t-t'$, time 
translational invariance (TTI) holds 
\begin{eqnarray} 
  && \lim_{t\rightarrow\infty} C_k(t,t) = \tilde q_k,      \\ 
  &&  \lim_{t\rightarrow\infty} C_k(t+\tau,t) = C_k(\tau), \\ 
  && \lim_{\tau\rightarrow\infty} C_k(\tau) = q_k,   
\end{eqnarray} 
and 
\begin{equation} 
  \lim_{t\rightarrow\infty} R_k(t+\tau,t) = R_k(\tau) . 
\end{equation} 
In addition to the TTI regime, there is another long time non trivial 
regime, characterized by $t,t'\rightarrow\infty$, fixing 
$\lambda=h(t')/h(t)$ and $0<\lambda<1$, where the function $h(t)$ is 
an increasing function of $t$ which the asymptotic analysis performed 
here is not able to determine. 
In this aging regime one has 
\begin{eqnarray} 
  && \lim_{t\rightarrow\infty} C_k(t,h^{-1} 
       (\lambda h(t))) = q_k \hat C_k(\lambda), \\ 
  && \lim_{\lambda\rightarrow 0} q_k \hat C_k(\lambda) = q_{0,k},        \\ 
  && \lim_{\lambda\rightarrow 1} \hat C_k(\lambda) = 1, 
\end{eqnarray} 
and 
\begin{equation} 
  \lim_{t\rightarrow\infty} R_k(t,\lambda t) = \frac{1}{t}\hat R_k(\lambda). 
\end{equation} 
Also, for future convenience, 
it is useful to introduce the following order 
parameters: 
\begin{eqnarray} 
 && \tilde q \equiv \lim_{t\rightarrow\infty}  
     \langle\langle x(s,t)x(s,t) \rangle\rangle  
    = \int \frac{dk}{2\pi} \ \tilde q_k,  \\ 
 && q \equiv \lim_{\tau\rightarrow\infty} \lim_{t\rightarrow\infty} 
    \langle\langle x(s,t+\tau)x(s,t) \rangle\rangle =  
    \int \frac{dk}{2\pi} \ q_k,  \\  
 && q_0 \equiv \lim_{\lambda\rightarrow 0} \lim_{t\rightarrow\infty} 
    \langle\langle x(s,t)x(s,\lambda t) \rangle\rangle  =  
    \int \frac{dk}{2\pi} \ q_{0,k},  
\end{eqnarray} 
together with 
\begin{equation} 
  b = 2 ( \tilde q - q ) \ , \ \ b_0 = 2 ( \tilde q - q_0 ).  
\end{equation}

\section{Equations relating asymptotic values of correlation  
and response functions} 
\label{sec:qs} 
 
Using the ansatz discussed in section \ref{sec:ansatz} one can derive 
the following equations for $C_k(t,t')$ in the TTI regime: 
\begin{eqnarray} 
  & \left[ T(\mu+k^2)+\partial/\partial\tau \right]  C_k(\tau) = &  
   2 T R_k(-\tau)  
  + \frac{2}{T} {\cal V}'(b) \left[ C_k(\tau) - q_k \right]   
  - \frac{2}{T} \int_{0}^{\tau} d\tau' {\cal V}'(B(\tau-\tau'))  
        \frac{\partial C_k(\tau')}{\partial\tau'} 
 \nonumber \\ 
 & &  +  2 \int_{0}^{1} d\rho {\cal V}'(\hat B(\rho)) \hat R_k(\rho)  
      +  4 \int_{0}^{1} d\rho {\cal V}''(\hat B(\rho)) \hat r(\rho)  
        \left[ C_k(\tau) - q_k \hat C_k(\rho) \right]  
  \label{eq:CkTTI} 
\end{eqnarray} 
It is also possible to derive similar equations for $R_k(\tau)$ which, 
due to the Fluctuation Dissipation Theorem (FDT) 
\begin{equation} 
  R_k(\tau) = - \frac{1}{T} \frac{d\,C_k(\tau)}{d\,\tau} 
  \label{eq:FDT}, 
\end{equation} 
are completely equivalent to Eq.~(\ref{eq:CkTTI}). 
 
In the aging regime one gets the following equation for $q_k\hat 
C(\lambda)$: 
\begin{eqnarray} 
  & \left[ T(\mu+k^2) \right. &  
    \left. -   4 \int_{0}^{1} d\rho {\cal V}''(\hat B(\rho)) \hat r(\rho)  
    \right]  q_k \hat C_k(\lambda)  =   
    2 \int_{0}^{1} d\rho {\cal V}'(\hat B(\rho)) \hat R_k(\rho)  
    +  \frac{2}{T} {\cal V}'(\hat B(\lambda)) ( \tilde q_k - q_k )  
  \nonumber \\ 
  && -  4 \int_{0}^{\lambda} d\rho {\cal V}''(\hat B(\rho)) \hat r(\rho)     
        q_k \hat C_k(\rho/\lambda)  
     -  4 \int_{\lambda}^{1} d\rho {\cal V}''(\hat B(\rho)) \hat r(\rho)     
        q_k \hat C_k(\lambda/\rho).  
  \label{eq:CkAG} 
\end{eqnarray} 
For $\hat R_k(\lambda)$ we obtain, 
\begin{eqnarray} 
   \left[  
      T(\mu+k^2)  
      -  4 \int_{0}^{1} d\rho {\cal V}''(\hat B(\rho)) \hat r(\rho)  
   \right] \hat R_k(\lambda)  = 
    - \frac{4}{T} {\cal V}''(\hat B(\lambda)) \hat r(\lambda)  
      ( \tilde q_k - q_k )  
     -  4 \int_{\lambda}^{1} \frac{d\rho}{\rho} {\cal V}''(\hat B(\rho))     
        \hat r(\rho) \hat R_k(\lambda/\rho) . 
  \label{eq:RkAG} 
\end{eqnarray} 
Again, one can see that both Eq.~(\ref{eq:CkAG}) and 
Eq.~(\ref{eq:RkAG}) can be solved by the ansatz 
\begin{equation} 
  \hat R_k(\lambda) = \frac{x}{T} q_k \frac{d\,\hat C_k(\lambda)}{d\,\lambda}.  
  \label{eq:GFDT} 
\end{equation} 
Eq.~(\ref{eq:GFDT}) is commonly referred to as a generalized FDT 
(GFDT). In principle, Eq.~(\ref{eq:GFDT}) could have been written as 
\begin{equation} 
  \hat R_k(\lambda) = \frac{ x_k(q_k\hat C_k(\lambda)) }{ T } \,  
                     q_k \frac{ d\hat C_k(\lambda) }{ d\lambda }, 
\end{equation} 
which could be applied to a many-step RSB scheme. However, as 
previously discussed, the present random heteropolymer model can be 
identified with the random manifold problem with short range potential 
correlations. As such, it has one step RSB, and it is sufficient to 
use the simpler ansatz given in Eq.~(\ref{eq:GFDT}). 
 
For $t=t'$ and $t\rightarrow\infty$ Eq.(\ref{eq:emC}) gives 
\begin{eqnarray} 
  T(\mu+k^2) \tilde q_k =  
     T + \frac{2}{T} {\cal V}'(b) ( \tilde q_k - q_k )  
     + 2 \int_{0}^{1} d\rho {\cal V}'(\hat B(\rho)) \hat R_k(\rho)  
     + 4 \int_{0}^{1} d\rho {\cal V}''(\hat B(\rho)) \hat r(\rho)  
          \left[ \tilde q_k - q_k C_k(\rho) \right] . 
  \label{eq:qtk1}  
\end{eqnarray} 
Eq.~(\ref{eq:CkTTI}) for $t\rightarrow\infty$ and then 
$\tau\rightarrow\infty$ results in 
\begin{eqnarray} 
  T(\mu+k^2) q_k = 
     \frac{2}{T} {\cal V}'(b) ( \tilde q_k - q_k )  
     + 2 \int_{0}^{1} d\rho {\cal V}'(\hat B(\rho)) \hat R_k(\rho)  
     + 4 \int_{0}^{1} d\rho {\cal V}''(\hat B(\rho)) \hat r(\rho)  
          q_k \left[ 1 - C_k(\rho) \right] . 
  \label{eq:qk1}  
\end{eqnarray} 
Also, Eq.(\ref{eq:CkAG}) for $\lambda\rightarrow 0$ gives 
\begin{eqnarray} 
  T(\mu+k^2) q_{0,k} =  
     2 {\cal V}'(b_0) \int_{0}^{1} d\rho \hat R_k(\rho) +  
     \frac{2}{T} {\cal V}'(b_0) ( \tilde q_k - q_k ) . 
  \label{eq:q0k1}   
\end{eqnarray} 

Eqs.~(\ref{eq:qtk1}), (\ref{eq:qk1}) and (\ref{eq:q0k1}) (and, 
equivalently, \ref{eq:CkTTI}, \ref{eq:CkAG} and \ref{eq:RkAG}) contain 
TTI and aging parts.  Thus, in principle, there are two ansätze for 
solving them, leading to two phases: an ergodic one (without aging) 
and a glassy one (with aging).

\section{Ergodic Phase}

Technically, assuming that aging is absent amounts to setting $\hat 
R_k(\lambda)=0$ and $\hat C_k(\lambda)=1$ in (\ref{eq:qtk1}), 
(\ref{eq:qk1}) and (\ref{eq:q0k1}). (Equivalently, one could start 
from (\ref{eq:emC}) and (\ref{eq:emR}) and exclude the aging part from 
the beginning, leading to the same equations.)  Thus, in the ergodic 
phase, equations (\ref{eq:qtk1}), (\ref{eq:qk1}) and (\ref{eq:q0k1}) 
reduce to 
\begin{eqnarray} 
   && T(\mu+k^2) \tilde q_k =  
      T + \frac{2}{T} {\cal V}'(b) (\tilde q_k - q_k),  
   \label{eq:qter} \\ 
   && T(\mu+k^2) q_k =  
      \frac{2}{T} {\cal V}'(b) (\tilde q_k - q_k) , 
   \label{eq:qker} \\ 
   && T(\mu+k^2) q_{0,k} =  
      \frac{2}{T} {\cal V}'(b_0) (\tilde q_k - q_k) . 
   \label{eq:q0ker} 
\end{eqnarray} 
Note that (\ref{eq:qker}) and (\ref{eq:q0ker}) enforce $q_k=q_{0,k}$ 
which is just equivalent to $\hat C_k(\lambda)=1$, so one gets only 
two equations. Solving them for $\tilde q_k$ and $q_k$ gives 
\begin{eqnarray} 
   && \tilde q_k - q_k = \frac{1}{\mu+k^2} \label{eq:bker} \\ 
   && \tilde q_k = \frac{1}{\mu+k^2} + \frac{2}{T^2} {\cal V}'(b)  
                   \frac{1}{(\mu+k^2)^2}. 
\end{eqnarray} 
After integrating over $k$ and using 
\begin{equation} 
  \int \frac{dk}{2\pi} \frac{1}{\mu+k^2}=\frac{1}{2\sqrt{\mu}} \ , \ \      
  \int \frac{dk}{2\pi} \frac{1}{(\mu+k^2)^2}=\frac{1}{4\mu^{3/2}}, 
  \label{eq:intk} 
\end{equation}  
we obtain 
\begin{eqnarray} 
  && q = \frac{1}{2\mu^{3/2}T^2} {\cal V}'(1/\sqrt{\mu}) , 
  \label{eq:qerg} \\ 
  && \tilde q = \frac{1}{2\sqrt{\mu}} +  
                \frac{1}{2\mu^{3/2}T^2} {\cal V}'(1/\sqrt{\mu}). 
  \label{eq:qterg} 
\end{eqnarray} 
For $T$ very small $q$ and $\tilde q$ blow up since the confinement 
term $\mu x(s,t)^2$ term becomes ineffective (see Eq.~\ref{eq:Z}). 
For very large temperature $q$ approaches zero but is never exactly 
equal to zero.

\section{Spin glass phase} 
 
Keeping the aging parts and using the GFDT, Eqs.~(\ref{eq:qtk1}), 
(\ref{eq:qk1}) and (\ref{eq:q0k1}) can be transformed into 
\begin{eqnarray} 
   && T(\mu+k^2) \tilde q_k = T   
      + \frac{2}{T} {\cal V}'(b) (1-x) ( \tilde q_k - q_k )  
      + \frac{2}{T} {\cal V}'(b_0) x ( \tilde q_k - q_{0,k} ) , 
   \label{eq:qtk2} \\ 
   && T(\mu+k^2) q_k = 
      \frac{2}{T} ( {\cal V}'(b) - x {\cal V}'(b_0) ) ( \tilde q_k - q_k ) 
      + \frac{2}{T} {\cal V}'(b_0) x ( \tilde q_k - q_{0,k} ) ,    
   \label{eq:qk2} \\ 
   && T(\mu+k^2) q_{0,k} = 
      \frac{2}{T} {\cal V}'(b_0) (1-x) ( \tilde q_k - q_k )   
      + \frac{2}{T} {\cal V}'(b_0) x ( \tilde q_k - q_{0,k} ) .    
   \label{eq:q0k2} 
\end{eqnarray} 
Solving Eqs.~(\ref{eq:qtk2}), (\ref{eq:qk2}) and (\ref{eq:q0k2}) for 
$\tilde q_k$, $q_k$ and $q_{0,k}$ gives 
\begin{eqnarray} 
   && \tilde q_k - q_k = \frac{ 1 }{ \mu + k^2 + \Sigma }, \label{eq:bk} \\ 
   && \tilde q_k - q_{0,k} =  
      \frac{1}{x} \frac{1}{\mu+k^2}  
      - \frac{1-x}{x} \frac{1}{\mu+k^2+\Sigma}, \\ 
   && \tilde q_k = ( \tilde q_k - q_{0,k} )  
      + \frac{2}{T^2} {\cal V}'(b_0) \frac{1}{(\mu+k^2)^2}, 
\end{eqnarray} 
where 
\begin{equation} 
   \Sigma = x \frac{2}{T^2} ( {\cal V}'(b) - {\cal V}'(b_0) ). 
\end{equation} 
Integration over $k$ and using (\ref{eq:intk}) 
gives 
\begin{eqnarray} 
  && b = \frac{1}{\sqrt{\mu+\Sigma}},  
  \label{eq:b} \\ 
  && b_0 = \frac{1}{x} \frac{1}{\sqrt{\mu}}  
           - \frac{1-x}{x} \frac{1}{\sqrt{\mu+\Sigma}} , 
  \label{eq:b0} \\ 
  && \tilde q = b_0 + \frac{1}{2\mu^{3/2}T^2} {\cal V}'(b_0) .        
  \label{eq:qt}    
\end{eqnarray} 
Furthermore, Eq.~(\ref{eq:RkAG}) with $\lambda=1$ gives 
\begin{equation} 
  \hat R_k(1) (\mu+k^2+\Sigma)  = - (\tilde q_k - q_k )  
  \frac{4{\cal V}''(b)}{T^2} \hat r(1), 
\end{equation} 
and after using Eq.~(\ref{eq:bk}), integrating over $k$ and using 
$\mu+\Sigma=b^{-2}$ (see Eq.~\ref{eq:b}) one gets 
\begin{equation} 
  0 = \hat r(1) \left[ T^2 + b^3 {\cal V}''(b) \right]. 
  \label{eq:hatr} 
\end{equation} 
Eq.~(\ref{eq:hatr}) with $\hat r(1)\ne 0$ implies 
marginal stability condition 
\begin{equation} 
  - T^2 = b^3 {\cal V}''(b). 
  \label{eq:bT}  
\end{equation} 
Also, equations (\ref{eq:b}) and (\ref{eq:b0}) can be rewritten as 
\begin{eqnarray} 
  && \frac{ {\cal V}'(b) - {\cal V}'(b_0) }{ b_0 - b } = \frac{T^2}{2} 
      \frac{\sqrt\mu}{b} \left( \frac{1}{b} + \sqrt{\mu} \right),  
  \label{eq:b0b} \\ 
  && b_0 - b = \frac{1}{x} \left( \frac{1}{\sqrt{\mu}} - b \right). 
  \label{eq:xbb0} 
\end{eqnarray} 
Eqs.~(\ref{eq:bT}), (\ref{eq:b0b}) and (\ref{eq:xbb0}) fully solve the 
model: (\ref{eq:bT}) gives $b$ as function of $T$, (\ref{eq:b0b}) 
determines $b_0$ as function of $T$ and $\mu$, (\ref{eq:xbb0}) 
determines $x(T,\mu)$ and Eq.~(\ref{eq:qt}) gives $\tilde 
q(T,\mu)$. Knowing $b(T)$, $b_0(T,\mu)$ and $\tilde q(T,\mu)$ 
determines $q(T,\mu)$ and $q_0(T,\mu)$.  Were we to impose the 
spherical constraint $\tilde q=const$, Eq.~{\ref{eq:qt}} could be used 
to relate $\mu$ and $T$, and all order parameters could be expressed 
as functions of $T$ only ($\tilde q$ being fixed). However, in this 
study we work with fixed $T$ and $\mu$ allowing $\tilde q$ to change.

\section{Solving the equations (Phase Diagram)} 
 
The procedure of solving equations similar to the ones given in 
Eqs.~(\ref{eq:bT}), (\ref{eq:b0b}) and (\ref{eq:xbb0}) has been 
discussed in ref. \cite{CD}.  We apply a similar analysis to the 
random heteropolymer problem.  In principle, three are three critical 
lines in the $T,\mu$ plane separating them (as shown in figure 1). 
 
{\bf Critical line (1):} $T=T_{max}$ is the uppermost critical line 
(denoted in figure 1 by (1)); above this line Eq.~(\ref{eq:bT}) has no 
solution.  The value of $T_{max}$ can be determined from the graphical 
solution of Eq.~(\ref{eq:bT}) depicted in figure 2.  Once $T$ has been 
chosen (horizontal line labeled $(T/T_{max})^2$) $b$ is found from the 
intercept of $(T/T_{max})^2$ line with $-b^3{\cal V}''(b)$ curve. 
From figure 2 it is clear that at $b=b_{max}$ the right hand side of 
Eq.~(\ref{eq:bT}) reaches a maximum; requiring $\frac{d}{db}\left[ 
b^3{\cal V}''(b) \right]=0$ gives $3{\cal V}''(b) + b{\cal V}'''(b)=0$ 
and $b_{max}=\frac{3\sigma}{\gamma-2}$. Accordingly, $T_{max}=\left[ 
-b_{max}^3{\cal V}''(b_{max}) \right]^{1/2}$. 
 
Also, note that for fixed $T$, Eq.~(\ref{eq:bT}) has two solutions for 
$b$ (denoted by $b_I$ and $b_{II}$ in figure 2).  The first, physical 
solution ($b_{I}\rightarrow 0$ for $T\rightarrow 0$), in the interval 
$[0,b_{max}]$ and second, unphysical solution 
($b_{II}\rightarrow\infty$ for $T\rightarrow 0$), in the interval 
$[b_{max},\infty)$.  Accordingly, a model with $\sigma=0$ (i.e.\ 
$V(\Delta x)=\delta(\Delta x)$) leads to an unphysical phase diagram, 
since for $\sigma\rightarrow 0$ physical branch $[0,b_{max}]$ 
disappears ($b_{max}\rightarrow 0$). 
 
Clearly, the form of ${\cal V}(b)$ for small $b$ has to be modeled 
carefully and the choice $V(\Delta x)=\delta(\Delta x)$ simply fails 
in that respect, giving ${\cal V}(0)=\infty$. Thus when formulating 
the problem, if there is to be a possibility of freezing at low 
temperatures ($b\rightarrow 0$ as $T\rightarrow 0$), the bead-bead 
interaction $V(\Delta x)$ has to be regular for small $\Delta 
x$. Similar small distance regularization problem of bead-bead 
interaction was encountered in ref \cite{PS}. 
 
{\bf Critical line (2):} corresponds to $b=b_0$. From 
Eq.~(\ref{eq:xbb0}) it follows that $b=b_0=1/\sqrt{\mu}$. The equation 
of the critical line is obtained by inserting $b=1/\sqrt{\mu}$ into 
(\ref{eq:bT}): 
\begin{equation} 
  (T/\tilde B)^2 =  
  \frac{\gamma}{2} \mu^{-3/2} (\mu^{-1/2}+\sigma )^{-(\gamma+1)}.   
  \label{eq:cl2} 
\end{equation} 
$\mu\in[\mu_{max},\infty)$, where $\mu_{max}$ solves Eq.~(\ref{eq:cl2}) 
with $T=T_{max}$.  
 
The value of $x_c$ at the critical line can not be directly obtained 
from Eq.~(\ref{eq:xbb0}). Instead, one has to approach the critical 
line and obtain the limiting value of x: for example, first, one 
assumes that point ($T_c$,$\mu_c$) is at the critical line ($T_c$ and 
$\mu_c$ satisfy Eq.~\ref{eq:cl2}) and, then, $T(\epsilon)$, 
$\mu(\epsilon)$, $b(\epsilon)$, $b_0(\epsilon)$, $x(\epsilon)$ 
approach their values at the critical line for $\epsilon\rightarrow 
0$. Naturally, the dependence on $\epsilon$ has to be chosen 
consistently with Eqs.~(\ref{eq:bT}), (\ref{eq:b0b}) and 
(\ref{eq:xbb0}). Since one has five variables and three equations 
which relate them, two variables have to be specified as, e.g., 
$b_0(\epsilon)=b_c+\epsilon$, with $b_c=1/\sqrt{\mu_c}$, and 
$T(\epsilon)=T_c$. The other three variables $b(\epsilon)$, 
$\mu(\epsilon)$ and $x(\epsilon)$ have to be determined from 
(\ref{eq:bT}), (\ref{eq:b0b}) and (\ref{eq:xbb0}): 
\begin{eqnarray} 
  && \frac{ {\cal V}'(b_c) - {\cal V}'(b_c+\epsilon) }{ \epsilon }  
      = \frac{T_c^2}{2} 
       \frac{\sqrt{\mu(\epsilon)}}{b_c}  
       \left( \frac{1}{b_c} + \sqrt{\mu(\epsilon)} \right),  
  \label{eq:b0bcl} \\ 
  && \epsilon = \frac{1}{x(\epsilon)}  
       \left( \frac{1}{\sqrt{\mu(\epsilon)}} - b_c \right). 
  \label{eq:xbb0cl} 
\end{eqnarray} 
Equation (\ref{eq:bT}) is trivially satisfied and does not enter the 
analysis.  At first order in $\epsilon$ Eqs.~(\ref{eq:xbb0cl}) and 
(\ref{eq:b0bcl}) give 
\begin{equation} 
  x(0) = - \frac{1}{2\mu_c^{3/2}} \mu'(0) \ , \ \ 
  \mu'(0) = - \frac{2}{3} \frac{{\cal V}'''(b_c)}{T_c^2\sqrt{\mu_c}}, 
\end{equation} 
which, together with $T_c^2=-\mu_c^{-3/2}{\cal V}''(b_c)$, gives the 
value for x at the critical line (2), 
\begin{equation} 
  x_c = - \frac{1}{3} \frac{{\cal V}'''(b_c)}{{\cal V}''(b_c)}. 
\end{equation} 
Using the explicit form for ${\cal V}$ gives 
\begin{equation} 
  x_c = \frac{\gamma+1}{3} \frac{b_c}{b_c+\sigma}. 
\end{equation} 
with $b_c$ on the critical line. $b_c=b_{max}$ gives 
$x_c=1$ while for $b_c=0$ one gets $x_c=0$. 
 
Thus, at the critical line (2), close to $T_{max}$, $x_c$ is very 
close to $1$ and as $T$ ($\mu$) decreases (increases) $x_c$ drops to 
zero.  Also, at the critical line (2), the transition to the ergodic 
phase is continuous in $b$ and $b_0$ and discontinuous in $x$. 
 
{\bf Critical line (3):} at this line $x=1$ and Eq.~(\ref{eq:xbb0}) 
gives $b_0=1/\sqrt{\mu}>b$. The equation for this critical line is 
given by (\ref{eq:b0bcl3}). 
\begin{equation} 
  \frac{ {\cal V}'(b) - {\cal V}'(1/\sqrt{\mu}) }{ 1/\sqrt{\mu} - b }  
   = \frac{T^2}{2} \frac{\sqrt\mu}{b} \left( \frac{1}{b} + \sqrt{\mu} \right).  
  \label{eq:b0bcl3} 
\end{equation} 
Once $T$ is chosen, $b$ is determined from (\ref{eq:bT}) and upon 
solving Eq.~(\ref{eq:b0bcl3}) one obtains $\mu$ as function of $T$. 
Critical line (3) is depicted in figure 1, where it was obtained by 
solving Eq.~(\ref{eq:b0bcl3}) numerically. The line starts from 
$(\mu_{max},T_{max})$ and then drops to $(0,T^*)$ where $T^*$ is given 
from Eq.~(\ref{eq:bT}) with $b=b^*$ and 
$b^*=\frac{2\sigma}{\gamma-2}$. Thus, as $b\rightarrow b^*$, 
$b_0\rightarrow\infty$, as can easily be checked by inserting those 
assumptions in Eq.~(\ref{eq:b0bcl3}). Also, $b_0\rightarrow b$ as 
$\mu\rightarrow\mu_{max}$.  Thus, contrary to (2), on line (3) the 
transition to the ergodic phase is discontinuous in $b$ and $b_0$ 
while continuous in $x$. 
 
Also, for arbitrary $\mu$, when $T$ gets close to zero $b$ approaches 
$0$ and $b_0$ grows to infinity. This simply means that for low 
temperatures the heteropolymer freezes completely: 
$x(s,t+\tau)=x(s,t)$ for arbitrary $\tau$ and $t$ sufficiently 
large. On the other hand, for fixed $T$ and vanishing $\mu$, 
Eq.~(\ref{eq:b0b}) gives $b_0\rightarrow\infty$, while $b$ stays fixed 
by Eq.~(\ref{eq:bT}). 
 
For small $\mu$ Eqs.~(\ref{eq:qerg},\ref{eq:qterg}) give $q/\tilde 
q\propto \mu^{(d-2)/4}$. Thus, for $\mu=0$ one gets $q/\tilde q=0$. 
Also, as discussed in the preceding paragraph, in the glass phase for 
$\mu \rightarrow 0$ one has $b_0\rightarrow\infty$ and $b= {\rm 
const}$, which gives $q/\tilde q=1$. Thus contrary to the ergodic 
phase, where vanishing $\mu$ lead to paramagnetic-like behavior, in 
the glass phase the system gets trapped in one of many states 
separated by diverging barriers.  Interestingly enough, adjusting 
$\mu$ so that the radius of gyration $R_g$ scales according to 
$R_g^d\sim N$ and using the relation $R_g^2 \sim 1/\sqrt{\mu}$ (which 
is exact for the Gaussian coil) \cite{TPW} gives $\mu\propto N^{-4/d}$ 
and $q/\tilde q\propto N^{-(d-2)/d}$. Thus, in the thermodynamic limit 
$q/\tilde q \rightarrow 0$.

\section{Discussion} 
 
We have presented a detailed derivation of the equations of motion of 
a random heteropolymer using SUSY formalism and a Gaussian variational 
ansatz.  In deriving these equations, we have used a long-chain 
approximation, considerably simplifying the dynamical 
action. Furthermore, imposing translational invariance, we have shown 
that, as happens in statics, within the Gaussian variational ansatz 
the equations of motion become identical to those for a manifold in a 
random potential with power law correlations. 
 
Clearly, this result is strongly related to the particular variational 
ansatz employed here, and its generality beyond this framework remains 
an open question.  Nevertheless, the existence of this mapping at the 
level of the GVA is rather intriguing.  It connects the random 
heteropolymer model with many physical systems, such as a manifold 
pinned by impurities, interfaces in a random field, the glassy phase 
of vortices in high-$T_c$ superconductors, directed polymers in a 
random potential, and surface growth on disordered substrates.  It 
would be interesting to understand to what extent the mappings to 
these problems extends beyond the GVA. 
 
By making the standards 1RSB aging ansatz for response and correlation 
functions we found the asymptotic solution of the dynamical 
equation. The validity of this ansatz has been carefully checked 
elsewhere: in the context of random manifold problem it was shown that 
one step replica symmetry breaking ansatz can be used to describe 
random manifold with short range correlations, and we have applied 
this results to the random heteropolymer. 
 
The analytic solutions show that, as expected, the random 
heteropolymer has characteristic properties of spin glass systems: 
aging and ergodicity breaking. Furthermore, the dynamical phase 
diagram is different from that for statics. In dynamics starting from 
a random condition, the polymer get stuck at energies higher then the 
ones of the native state. 
 
In a more realistic approach to heteropolymers, we expect that finite 
dimensional, and finite length chain effects will be responsible for 
ultimate restoring of ergodicity. Our study should be taken as an 
indication of a time regime where trapping effect and aging could be 
observed. 
 
One of the motivations for this paper, mentioned at the beginning of
the introduction, was the hope that it might provide some insight into
the dynamics of proteins, including their folding.  However, it is
fairly well understood by now that protein dynamics are influenced
strongly by the existence of an energetically favored native state, a
feature absent from the random heteropolymer model we have studied
here.  In work currently in progress, we are extending the analysis
presented here to models in which the two-body interactions $B_{s,s'}$
are systematically biased, with a tunable strength, to favor
particular ``native'' states.  Such models provide an opportunity to
study the competition between the attraction to a native state and the
glassiness produced by the randomness and frustration.
(Refs.~\cite{RamShak,WildShak,PGTbiophysj,PGTRMP} treat equilibrium
aspects of this competition.)

\acknowledgements

It is a pleasure to thank S. Solla for useful interactions at an early  
stage of this work.

\begin{figure} 
\epsfxsize=9cm 
\epsfysize=8cm 
\epsfbox{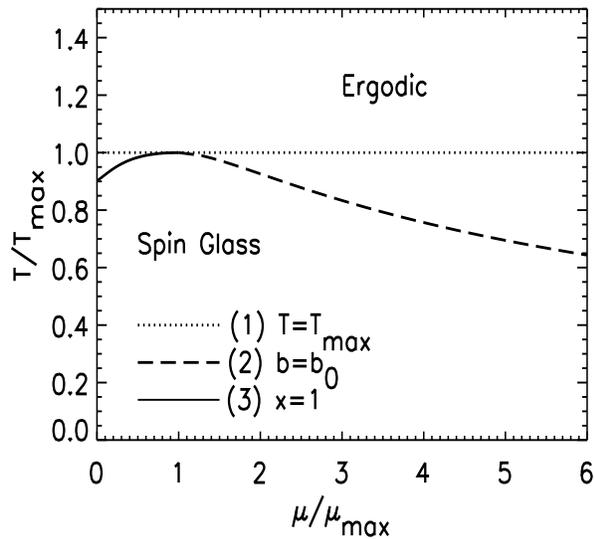} 
\caption{Phase diagram of dynamic random heteropolymer model in $\mu$, 
$T$ plane.  Critical lines are denoted by (1) $T=T_{max}$; (2) 
$b=b_0=\mu^{-1/2}$ ($x$ ranges from 1 to 0); (3) $x=1$, 
$b_0=\mu^{-1/2}>b$. Below (2) and (3) lies glassy phase and above 
ergodic phase} 
\end{figure} 
 
\begin{figure} 
\epsfxsize=9cm 
\epsfysize=8cm 
\epsfbox{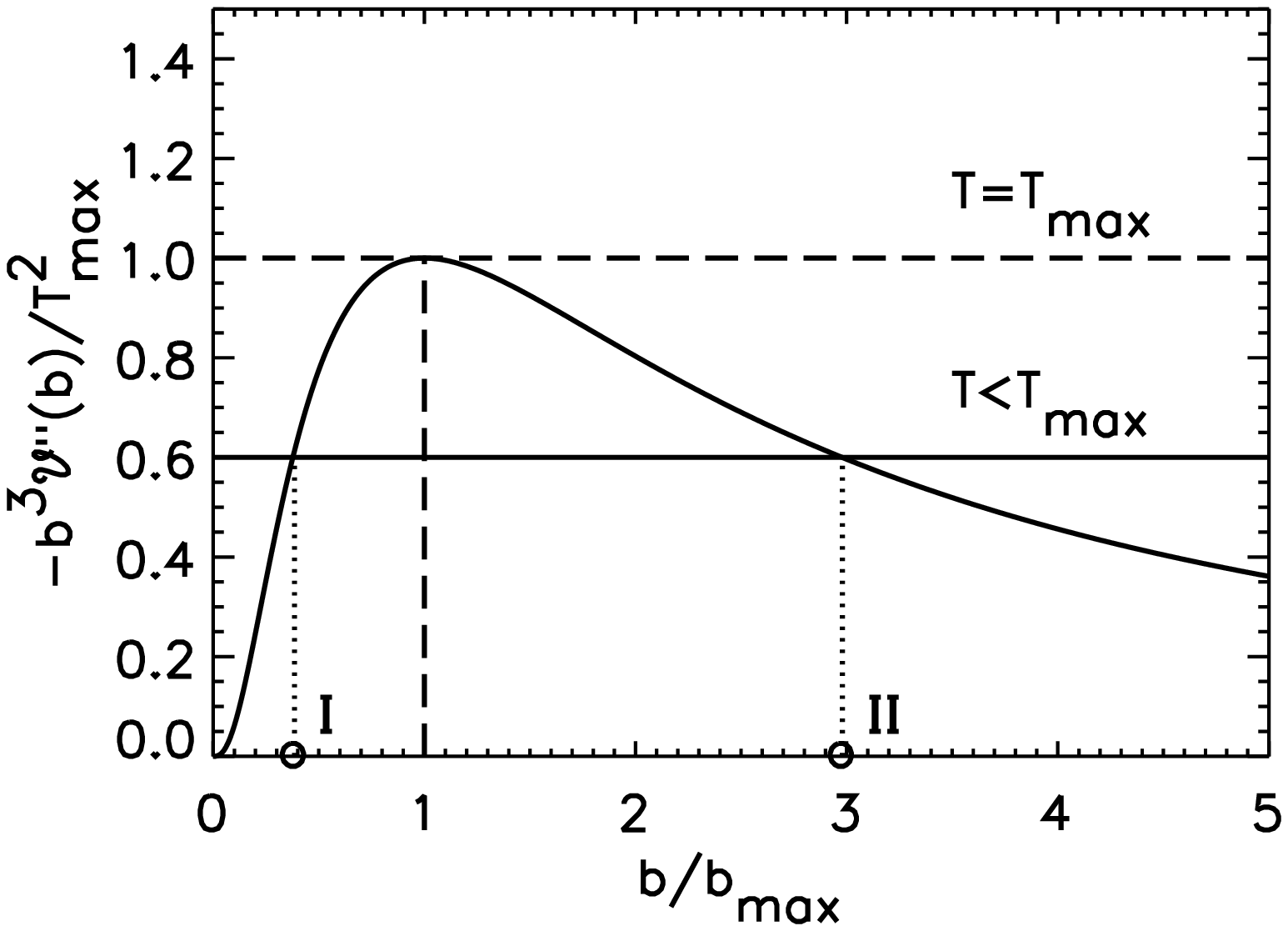} 
\caption{Graphical solution of Eq.~(\ref{eq:bT}). Equation has two 
solutions for $T<T_{max}$ denoted by ``I'' and ``II''. For $T=T_{max}$ 
there is only one solution $b=b_{bmax}$. Solution I is physical and 
solution II is unphysical.} 
\end{figure}


\begin{thebibliography}{19} 
 
 
\bibitem{SpGl} M. Mezard, G. Parisi and M.A. Virasoro, {\em Spin Glass 
Theory and Beyond} (World Scientific, 1987). 
 
\bibitem{WE} P.G. Wolynes and W.A. Eaton, Physics World {\bf 9}, 39 
(1999). 
 
\bibitem{NR} T. Nattermann and P. Rujan, Int. J. Mod. Phys. {\bf B 3}, 
1597 (1989). 
 
\bibitem{Wol1} P.G. Wolynes, H. Frauenfelder and R.H. Austin,  
{\em More things in heaven and earth. A celebration of physics at 
the millennium}, (Springer, 1999), page 706-725. 
 
 
 
\bibitem{Wol2} J.D. Bryngelson and P.G. Wolynes, 
Proc. Natl. Acad. Sci. USA {\bf 84}, 7524 (1987). 
 
 
 
\bibitem{garel} T. Garel and H. Orland, Europhys. Lett. {\bf 6}, 307 (1988) 
 
\bibitem{SG1} E. I. Shakhnovich and A. M. Gutin, 
Europhys. Lett. {\bf 8}, 327 (1989). 
 
\bibitem{SG2} E. I. Shakhnovich and A. M. Gutin, 
J. Phys. A {\bf 22}, 1647 (1989). 
 
\bibitem{GHLO} T. Garel, D.A. Huse, L. Leibler, H. Orland, 
Europhys. Lett. {\bf 8}, 9 (1989). 
 
\bibitem{SW} M. Sasai and P.G. Wolynes, Phys. Rev. Lett. {\bf 65}, 
2740 (1990). 
 
\bibitem{GLO} T. Garel, L. Leibler and H. Orland, J. Phys. II (France) 
{\bf 4}, 2139 (1994). 
 
\bibitem{TW} S. Takada and P.G. Wolynes, Phys. Rev. E {\bf 55}, 4562 
(1997). 
 
\bibitem{GOP} T. Garel, H. Orland and E. Pitard, cond-mat/9706125 
 
 
 
 
\bibitem{RS} J.R. Roan and E.I. Shakhnovich, Phys. Rev. E {\bf 54}, 
5340 (1996). 
 
\bibitem{TAB} D. Thirumalai, V. Ashwin and J.K.  Bhattacharjee, 
Phys. Rev. Lett. {\bf 77}, 5385-5388 (1996). 
 
\bibitem{TPW} S. Takada, J.J. Portman and P.G. Wolynes, 
Proc. Natl. Acad. Sci. USA {\bf 94}, 2318 (1997). 
 
\bibitem{Olem1} A.I.Olemskoi, Physica A {\bf 270}, 444-452 (1999). 
 
\bibitem{Pit} E. Pitard, Eur. Phys. J B {\bf 7}, 665-673 (1999). 
 
\bibitem{LT} N. Lee and D. Thirumalai, J. Chem. Phys. {\bf 113}, 
5126-5129 (2000). 
 
\bibitem{Olem2} A.I.Olemskoi, V.A.Brazhnyi, Physics of the Solid State 
{\bf 43}, 386-396 (2001). 
 
\bibitem{PS} E. Pitard and E.I. Shakhnovich, Phys. Rev. E {\bf 63}, 
041501 (2001). 
 
 
 
 
 
\bibitem{elba} S. Franz, M. Mezard and G. Parisi Int. J. Neural Systems,   
{\bf 3}, 195 (Supp. 1992)  
 
 
 
\bibitem{MP1} M. M\'{e}zard and G. Parisi, J. Phys. I {\bf 1}, 809 (1991). 
 
\bibitem{MP2} M. M\'{e}zard and G. Parisi, J. Phys. I {\bf 2}, 2231 
(1992). 
 
 
 
\bibitem{CD} L.F. Cugliandolo and P. Le Doussal, Phys. Rev. E {\bf 
53}, 1525 (1996). 
 
\bibitem{CKD} L.F. Cugliandolo, J. Kurchan and P. Le Doussal, 
Phys. Rev. Lett. {\bf 76}, 2390 (1996). 
 
 
\bibitem{MSR} P.C. Martin, E.D. Siggia and H.A. Rose, Phys. Rev. A 
{\bf 8}, 423 (1973);  
 
\bibitem{Dom} C. De Dominicis, Phys. Rev. B {\bf 18}, 4913 (1978). 
 
\bibitem{Kur} J. Kurchan, J. Phys. I (France) {\bf 2}, 1333 (1992). 
 
\bibitem{FM} S. Franz, M.~M\'ezard, { Europhys. Lett.} {\bf 26} (3) 
(1994) 209 and { Physica A } {\bf 210} (1994) 48. 
 
\bibitem{Eng} A. Engel, Nucl. Phys. B {\bf 410}, 617 (1993). 
 
\bibitem{KH1} H. Kinzelbach and H. Horner, J. Phys. I (France) {\bf 
3}, 1329 (1993). 
 
\bibitem{KH2} H. Kinzelbach and H. Horner, J. Phys. I (France) {\bf 
3}, 1329 (1993). 
 
 
\bibitem{CK1} L.F. Cugliandolo and J. Kurchan, J. Phys. A {\bf 27}, 
5749 (1994) 
 
\bibitem{BCKP} A. Baldassarri, L.F. Cugliandolo, J. Kurchan and 
G. Parisi, J. Phys. A {\bf 27}, 5749 (1994) 
 
\bibitem{CK2} L.F. Cugliandolo and J. Kurchan, Phil. Mag. B {\bf 
71}, 501 (1995). 

 
\bibitem{RamShak} S. Ramanathan and E. Shakhnovich,  Phys. Rev. E 
{\bf 50}, 1303 (1994). 
 
\bibitem{WildShak} J. Wilder and E. Shakhnovich, Phys. Rev. E {\bf 62}, 
7100 (2000). 
 
\bibitem{PGTbiophysj} V.S. Pande, A.Yu. Grosberg and T. Tanaka, 
Biophys. J. {\bf 73}, 3192 (1997) 
 
\bibitem{PGTRMP}  V.S. Pande, A.Yu. Grosberg and T. Tanaka,  
Rev. Mod. Phys. 72, 259 (2000). 
 
 
\end{thebibliography}
\end{document}